# Adaptive Inflow Control System

**Vasily Y. Volkov**

**Alexander P. Skibin,** PhD

Bauman Moscow Technical State University,

ul. Baumanskaya 2-ya, 5/1, Moscow, 105005 Russia

**Oleg N. Zhuravlev,** PhD

**Marat T. Nukhaev,** PhD

**Roman V. Shchelushkin**, PhD

WORMHOLES Ltd.,

ul. Ozerkovskaya naberezhnaya, 50 str. 1, Moscow, 115054 Russia

**Abstract.** This article presents the idea and realization for the unique Adaptive Inflow Control System being a part of well completion, able to adjust to the changing in time production conditions. This system allows to limit the flow rate from each interval at a certain level, which solves the problem of water and gas breakthroughs.

We present the results of laboratory tests and numerical calculations obtaining the characteristics of the experimental setup with dual-in-position valves as parts of adaptive inflow control system, depending on the operating conditions. The flow distribution in the system was also studied with the help of three-dimensional computer model. The control ranges dependences are determined, an influence of the individual elements on the entire system is revealed.

### Introduction

Currently, the oil and gas industry has two main types of inflow control systems with well completion. The commonly used inflow control systems are passive devices installed in sand screens. Nozzle or pipe/channel system modifications are the most widespread ones. These systems use hydraulic impedance to develop a certain pressure drop between the formation and the well thereby changing the pressure drawdown upon the formation. The level of hydraulic impedance (set by selecting a certain diameter of the nozzle or the length and diameter of the pipe/channel system) is selected based on log data obtained after drilling or logging while drilling (LWD) data and obviously this parameter cannot be changed after equipment has been installed into the well.

The main disadvantage of such systems is the lack of capability of changing their settings if the parameters of the near wellbore zone formation change in time or limiting the inflow of water or gas in case of their breakthrough. If the system is installed incorrectly or lowered to an insufficient depth, the entire assembly will fail to match the liquid inflow along the horizontal well thus deleteriously affecting the well rate and the overall oil field production.

Also known are active systems with hydraulically controlled valves that are installed on the production pipe inside the completion or the sand screens. These valves allow controlling the throttling for each zone from the ground. The main disadvantages of such systems are the high cost of equipment and servicing works during installation and operation, limited lowering depth and low reliability of operation.

WORMHOLES LLC. is invented and patented the autonomous fluid flow adjustment device able to provide the adjustment of completion to the fluid inflow due to the specific design features.

The device comprises an input and an output openings for fluid passage between which there is at least one sequential adjustment stage consisting of sequential or parallel installed hydraulic impedance and valve and/or an insulating (cutoff) element comprising a valve or a series of parallel installed valves that are not capable of automatically opening upon a decrease in the flow of the fluid, and furthermore said device comprises an outer enclosure in which said device components, or at least their part, are contained.

The autonomous fluid flow adjustment device operates as follows. During equipment installing into the well all the valves are open. The liquid and/or gas flow passes through the input device and the open valves and is delivered to the production pipe.

When the flow rate through the valve becomes greater than a certain (preset) value, the valve closes. The fluid flow duct changes, and the flow in the adjustment stage is redirected to flow through an additional hydraulic impedance and the open valve of the next adjustment element. The overall hydraulic impedance of the device increases and the flow drops. If the flow increases again, the next valve in the other adjustment element closes and the flow will return to the hydraulic impedance, etc.

When the flow rate of the fluid reduces (to below the preset value) the gate of the valve of the adjustment element returns to the initial open position thus changing the liquid flow duct and a decrease in the overall hydraulic impedance of the device. The fluid flow rate increases again. If the flow rate decreases again, the next valve in the other adjustment element opens and the flow will bypass the hydraulic impedance of the adjustment element thus increasing the overall flow rate through the device, etc.

Thus, liquid flow rate through the autonomous flow adjustment device will be kept within the preset range regardless of inflow rate.

In some embodiments of the device the adjustment element may comprise a normally open valve and a throttle as hydraulic impedance that can be interconnected either in sequence or in parallel.

In different embodiments of the device, depending on the well completion equipment used, the throttle may be, among others, any known flow diverting and/or reverting, merging and/or splitting, narrowing and/or broadening devices as well as combinations thereof. The choice of a specific hydraulic impedance type depends on the expected operation conditions of the device.

In some embodiments of the device the adjustment stage may comprise a normally open valve and a nozzle as hydraulic impedance that can be interconnected either in sequence or in parallel. The valve and nozzle connection type depends on the well completion equipment used.

For example, the nozzle can be a device for producing a pressure drop by increasing flow speed in an opening (nozzle, pipe etc.).

The operation of the valve used in the autonomous flow control device is based on the interaction of two physical phenomena, i.e. magnetism (or electric magnetism) and hydrodynamics. For the normally open valve option said interaction is as follows. During media flow through the flow duct formed by the valve design components, the pressure due to the impact velocity of the media and the hydraulic impedance of the flowing media and the pressure reduction in high flow rate areas change the pressure at the gate surface. As a result the valve gate is loaded by the hydrodynamic force which is counteracted by the force of at least one magnet or electric magnet comprised by the valve. As the flow rate increases the hydrodynamic force loading the gate at the side of the flowing media becomes greater than the magnetic force, and the valve condition changes. The gate lowers to the saddle and the valve closes. In some embodiments of the device, the magnet or electric magnet of gate retaining component are located at least at one side of the gate. This simplifies the design of the valve.

In some embodiments of the device, the magnet or electric magnet of gate retaining component are located at more than one side of the gate.

One of the elements of this system is open dual-in-position direct acting valve with a spherical locking element with a permanent magnet. We have performed the experimental tests and numerical calculations for the valve and the whole system (Figure 1). The prototype valve consists of two main elements: the cover – 1 and

saddle – 2 (Fig. 1a). The ball of magnetic material works as a locking element 3. The locking element is pressed against the valve cover under the influence of coercive force of the permanent magnets 4 (Fig. 1b) inserted in the valve cover.

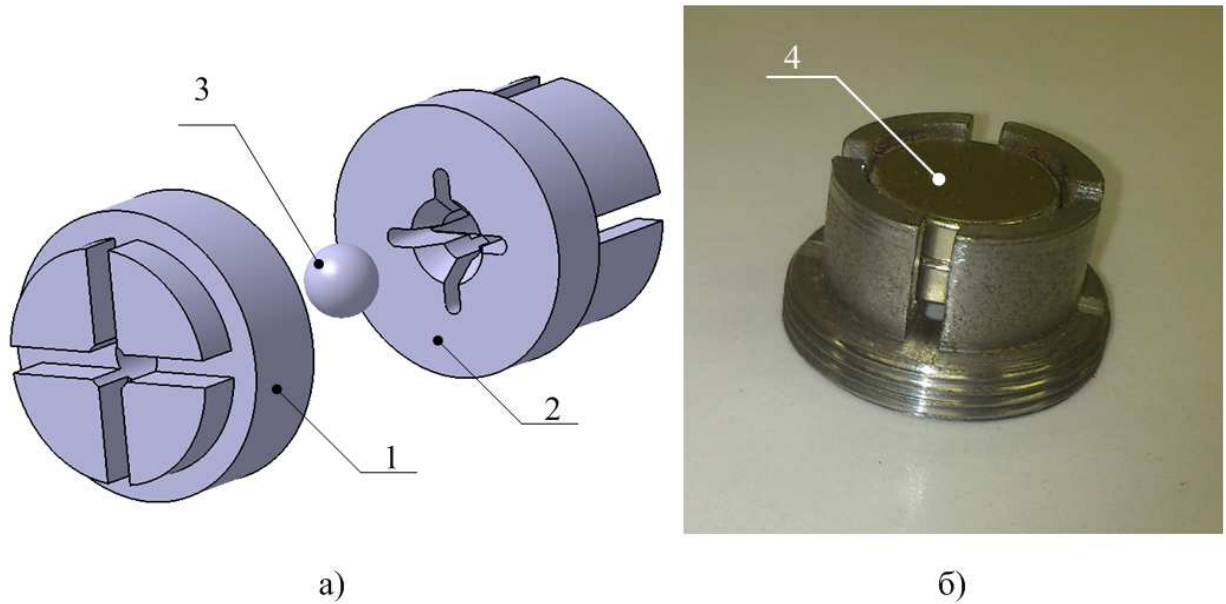

Figure 1. Valve design:

a) three-dimensional model of the valve; b) valve gate retaining component.

1 – valve saddle; 2 – valve cap; 3 – gate; 4 –magnets.

When the flow of liquid and/or gas through the AICD is less than nominal, the locking element (shutter) of the magnetic valve is open and held in a fixed position by the coercive force of the permanent magnet. In case of increasing the liquid flow rate higher than the nominal, the valve triggering occurs, the valve shutter closes and the flow is redirecting to the hydraulic resistance followed the valve, thereby increasing the overall hydraulic resistance of the system, which leads to a decrease in fluid flow rate in the case of constant pressure drop between the reservoir and the production pipe. Study of characteristics of throttle as a hydraulic resistance is presented in [1].

The experimental setup is developed at the Department of E-5 of Bauman Moscow Technical State University for laboratory testing of dual-in-position valves with magnets on the gas working medium (air). Experimental setup is designed to receive hydraulic flow characteristics on the pressure drop for a single valve and for a system consisting of three parallel valves. In this case parallel connection of the valves enables a continuously adjustable flow rate within a predetermined range. Ball valves were used as a hydraulic resistance in the experimental setup.

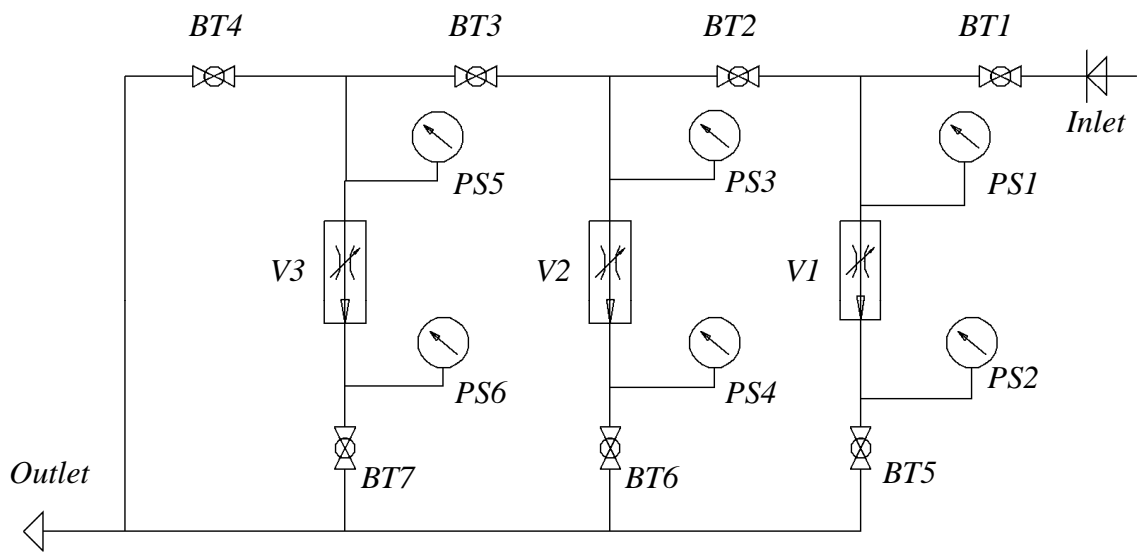

Figure 2. Pneumatic diagram of the experimental setup:

V1-3 – valves; PS1-6 – pressure sensors; BT1-7 – ball taps.

The experimental investigation has demonstrated a non-uniformity of valves triggering depending on the hydraulic resistance. This non-uniformity can be the cause of flow redistributions between the individual sections of the system.

A numerical simulation of the entire experimental setup with the use of computational aero- hydrodynamics (CFD – computational fluid dynamics) was conducted in addition to study of the system operation with fixed hydraulic resistors. CFD model of the experimental setup, which imitates a work of the setup for one position of hydraulic resistance of the main line, taking into account different modes with open and closed valves, was developed with the help of software package STAR-CCM +.

**Problem statement of numerical simulation**

In general, the problem statement includes a set of geometric model, the physical properties of the working environment and the mathematical model with the accompanying boundary conditions. Based on the design of the experimental setup a solid model of its flowing part was created in a graphics editor CATIA (see. Figure 3). The computational domain was divided into three characteristic regions: the main line with ball valves which simulating AICD hydraulic resistances, work areas with valves and the bypass line.

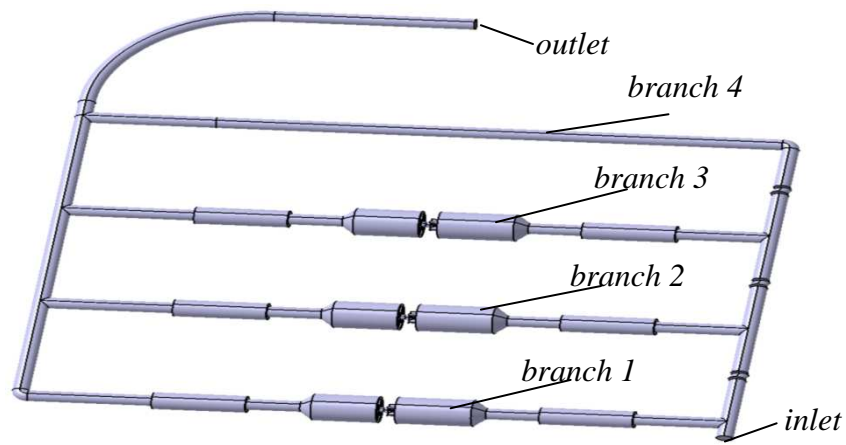

Figure 3 – Flowing part of the experimental setup

The built-in STAR-CCM+ grid generator of the polyhedral cells (Advanced Layer Mesher) [2] has used for the meshing of the computational domain. In the application of the polyhedral cells, the computational domain is divided into control volumes of complex polyhedral form which is close to sphere. This type of cells is the most economical in the automatic construction of a grid of the computational domain, as well as the most advantageous in terms of computational performance [2].

In the process of numerical calculations in accordance with the recommendations [3] the computational grid was refined and varied. A local thickening of the computational grid in the region of large gradients in the elements of the flow part was conducted. Figures 4a and 4b show computational grid of the flow part in the area of dual-in-position AICD valves before and after the local thickening respectively.

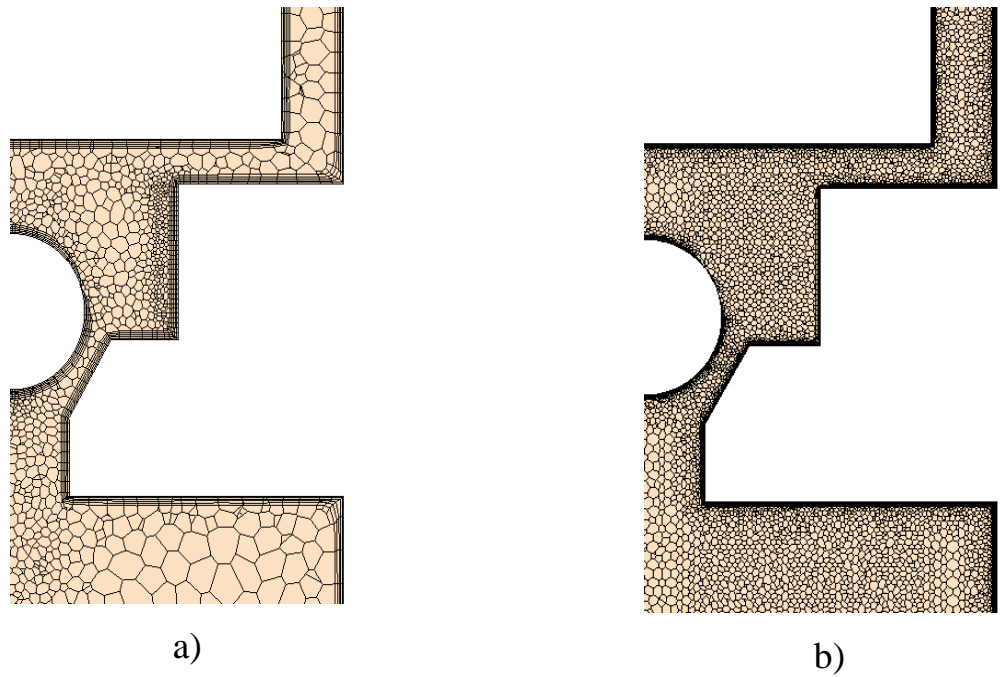

a)  b)

Figure 4 – Grid of the computational domain in the valve area: a) base; b) adapted

The following assumptions were made in the simulation of the flow of the working medium in the experimental setup:

- working medium (air) is considered as a Newtonian compressible fluid;
- the flow of the working medium is stationary;
- parameters of the working medium obeys the ideal gas law;
- the flow regime is turbulent;
- the flow of the working medium is adiabatic;
- viscosity, thermal conductivity and heat capacity of the medium are constant.

Given the extreme time-cost of methods based on the large eddy simulation models (LES) and direct numerical simulation (DNS), they are not suitable for rapid analysis of the characteristics of the experimental setup. Given this fact, to ensure efficiency analysis is necessary to use less costly approaches based on the Navier-Stokes equations averaged by Reynolds method.

The Reynolds equations must be supplemented for closure by the equations for the generation of turbulence $k$ and its dissipation $\varepsilon$ [4]. An approach to the closure of the Reynolds equations with nonlinear cubic modification of $k - \varepsilon$ model was used in the development of three-dimensional model of the experimental setup. Such variation of $k - \varepsilon$ model allows consider secondary flows with high accuracy, including channels of high curvature. However, additional nonlinear terms of the equations may lead to numerical instable solutions, especially on rough difference grids. To ensure the "physicality" of the results obtained using this model, the coefficients are determined taking into account the feasibility of the concept (Realizable [5]). This approach allows avoid the negative values of the variables $k$ and $\varepsilon$, as well as negative values of the Reynolds stresses.

The legitimacy of using the nonlinear cubic $k - \varepsilon$ turbulence model is validated by CFD verification study of individual elements of the setup. The system of equations describing the flow of the medium and used turbulence model is presented in [2]. The following boundary conditions were used for closure of the mathematical model:

The full pressure and temperature values are set at the entrance to the computational domain (G1, Figure 5) for the equations of motion and energy:

$$p = p_{in}$$

$$T = T_{in} = 300K$$

Wherein the temperature of the working medium was set constant for all models.

The inverse problem was solved to determine the pressure at the input when the value was determined from the condition that the pressure drop across the valve equals to 0.73 bar. The pressure was measured in a cross section in a place of installation of pressure sensors.

A constant pressure to the equations of motion is set at the outlet of the computational domain (G2, Figure 5):

$$p = p_{out}$$

Wherein the pressure at the outlet was assumed constant for all variants of the calculation - 1 bar.

Slip condition for the equations of motion and adiabatic condition for energy equation are set on the other boundaries (G3, Figure 5) of the computational domain:

$$u = 0; \quad \frac{\partial T}{\partial n} = 0$$

The computational domain was reconstructed in accordance with the position of the valves. The branch corresponding to the closed gate was removed from the computational domain when closing the valve.

Thus, a series of fixed simulations was carried out for different geometries corresponding to valves positions (closed/open) and variable boundary conditions to be determined from the condition of closing the gates of dual-in-position valves.

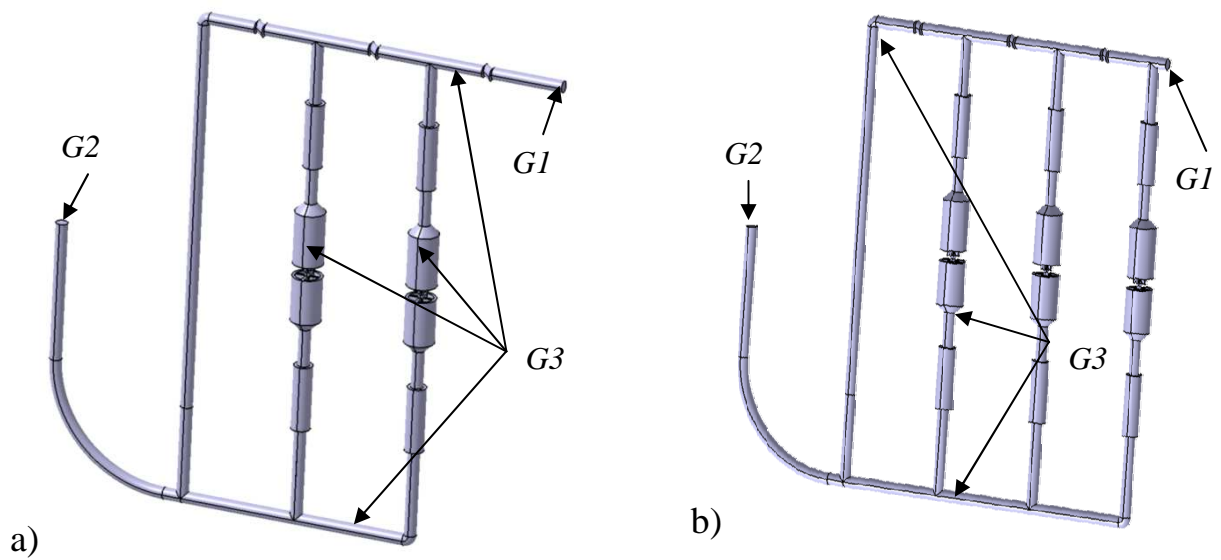

a) b)

Figure 5. Boundary conditions: a) to work with the two valves; b) to work with one valve

The computational domain in accordance with the scheme of the experimental setup (Figure 1, 3) is in four sections: three parallel branches with valves (in turn trigger) and the fourth branch of the hydraulic resistors line.

**The results of the numerical experiment**

Distribution of flow between the sections and the pressure drop in these parts are presented in the table. As can be seen from the analysis of the table, the flow rate through the first valve is only 25% of total flow rate and maximum flow rate (41%) is through the straight section 4. It is also worth noting that the ratio presented in the table will change at change in the resistance of ball valves of the main line.

Table – flow distribution into the branches and pressure drop across the valves

| Branch # | Flow rate $G_i$, kg/s | Relative flow rate $G_i/G_\Sigma \cdot 100\%$ | Pressure drop, $\cdot 10^5$ Pa |
|---|---|---|---|
| 1 | 0,69e-2 | 27,8 | 0,7394 |
| 2 | 0,48e-2 | 19,2 | 0,37 |
| 3 | 0,30e-2 | 12 | 0,15 |
| 4 | 1,03e-2 | 41 | |
| Σ | 2,515e-2 | 100 | |

Figure 6 shows the pressure field for different modes of AICD operation: while all valves are opened (Figure 6a), while working with one closed valve at different values of the inlet pressure (Figure 6b, 6c) and the mode of operation when one valve is open (Figure 6d). The data obtained were averaged and treated in accordance with the location of the sensors in the full-scale experimental study.

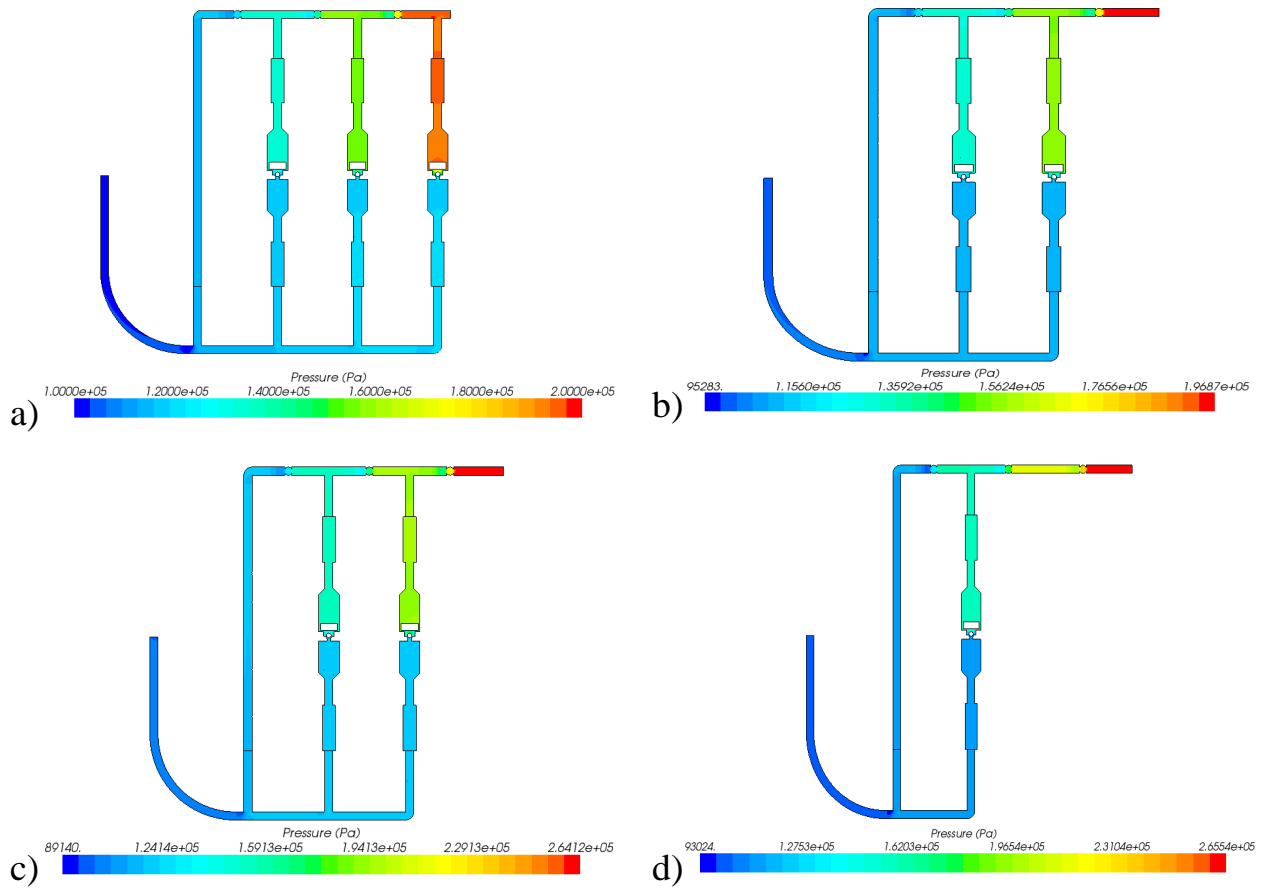

Figure 6. Pressure field for different modes of AICD operation:
a) all valves are opened ($p_{in}=2·10^5$ Pa); b,c) one valve is closed ($p_{in}=2·10^5$ Pa; $p_{in}=2,7·10^5$ Pa); d) one valve is opened ($p_{in}=2,7·10^5$ Pa)

Following the results of the simulation of gas dynamic setup its hydraulic characteristics of flow rate from pressure drop was obtained (Figure 7). As can be seen from the figure, the dependence shows that setting up the same resistance to the main line leads to increasing of flow rate, which is triggered by the dual-in-position valve, with subsequent triggering of the next valve. Wherein the third valve is activated out of control range to be set at 30% level. Thus, the determination of the hydraulic characteristics of the setup for the further setup operation is out of practical interest, since the activation of the third valve in the numerical experiment takes place outside the control range. Further operation of the system is determined only by the hydraulic resistance of the main line, which consists of three consecutive hydraulic resistances of ball valves, hydraulic resistance of friction and the local hydraulic resistance associated with the reversal of flow before exiting.

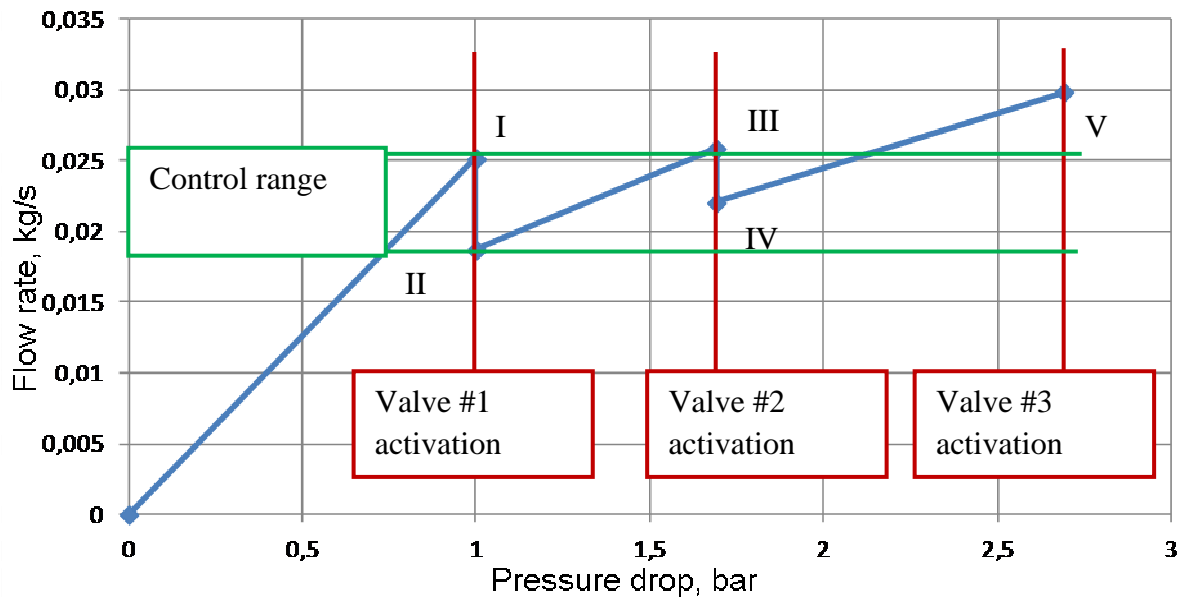

Figure 7 – Dependence of total flow rate from the pressure drop: I-V – calculations

Thus, the configuration of inflow control system in the specified range should make for various hydraulic resistances selected on the basis of hydraulic calculations. These resistances should be determined for the respective operating conditions of the system.

However, the computational study has shown that the ratio of the flow rates in the different areas is almost independent of the inlet pressure, and it is determined only by the hydraulic resistance of ball valves of the main line. Thus, system configuration can be made using less expensive techniques based on the balance equations for the contour networks [6], and CFD-models may thus be used to obtain closure ratios.

**Conclusion.**

The flow distribution through the branches of the experimental setup for open and closed valve positions was determined for the each position of direct action dual-in-position valves with the magnets.

The pressure drop values at the valves and the flow rates on branches for open and closed positions of the valves located in the experimental setup was determined. The calculations showed that there is the flow redistribution through the branches in accordance with the hydraulic resistance of the regions.

It was shown that the ratio of the flow rates through the regions is substantially independent of the inlet pressure, and it is determined only by the hydraulic resistance of ball taps of the main line. However, the flow rate through the regions is unevenly distributed before and after the closing of dual-in-position valves. Wherein it is shown that the setting of adaptive inflow control system must be carried out for different hydraulic resistances selected on the basis of hydraulic calculations. These resistances should be determined for the respective conditions of the system operation.

The main advantage of the adaptive inflow control system is its ability to adapt to changing conditions in the near-wellbore zone. Also, this system allows you to restrict the flow from the interval at a certain level, which solves the problem of water and gas breakthroughs. Thus, the adaptive inflow control system enables to adjust an optimal operation of the well with the help of completion system throughout the whole operating time.